\documentclass[preprint,showpacs,preprintnumbers,amsmath,amssymb]{revtex4}

\usepackage{graphicx}
\usepackage{dcolumn}
\usepackage{bm}
\usepackage{amsmath,dsfont}
\usepackage{amssymb,amsthm,amscd, amsbsy, array}

\newcommand{\half}{{\scriptstyle{\frac{1}{2}}}}
\def\2{{\half}}
\newcommand{\const}{\mathop{\rm const}\nolimits}

\def\dAlembert{\vcenter {
    \hbox {\vrule height8pt width0.4pt depth0.0pt
           \vrule height8pt width7.2pt depth-7.6pt
           \vrule height8pt width0.4pt depth0.0pt
           \kern-8pt
           \vrule height0.4pt width8pt depth0.0pt
          \,}}}

\def\p{{\partial}}

\def\vb{{\vec{b}}}

\def\beq{\begin{equation}}
\def\eeq{\end{equation}}
\def\beqa{\begin{eqnarray}}
\def\eeqa{\end{eqnarray}}

\def\barray{\left(\begin{array}}
\def\earray{\end{array}\right)}
\def\barraynb{\begin{array}}
\def\earraynb{\end{array}}

\def\SO{{\rm SO}}
\def\smallover#1/#2{\hbox{$\textstyle\frac{#1}{#2}$}} %

\def\vx{{\vec{x}}}
\def\vy{{\vec{y}}}
\def\vX{{\overrightarrow{X}}}
\def\vY{{\overrightarrow{Y}}}

\def\vXi{{\overrightarrow{\Xi}}}
\def\vUpsilon{{\overrightarrow{\Upsilon}}}
\def\vA{{\vec{A}}}
\def\va{{\vec{a}}}
\def\vnabla{{\overrightarrow{\nabla}}}
\def\vbeta{{\vec{\beta}}}

\begin{document}

\preprint{arXiv:1105.4401v2}

\title{Kohn's theorem and Galilean symmetry}

\author{P-M. Zhang}\email{zhpm-at-impcas.ac.cn}

\author{P.~A.~Horvathy\footnote{On leave from 
{\it Laboratoire de Math\'ematiques et de Physique
Th\'eorique},
 Universit\'e de Tours 
(France).}}\email{ horvathy-at-lmpt.univ-tours.fr}
\affiliation{Institute of Modern Physics, Chinese Academy of Sciences
\\
Lanzhou (China)
\\
}

\date{\today}

\begin{abstract}
The relation between the
 separability of a system of charged particles in a uniform magnetic field and
Galilean symmetry is revisited using Duval's ``Bargmann
framework''. If the charge-to-mass ratios of the particles are identical, $e_a/m_a=\epsilon$ for all particles,
then the Bargmann space of the  magnetic  system is isometric to that of an anisotropic harmonic oscillator.
Assuming that the particles interact through a potential
which only depends on their relative distances, 
the system splits into one representing the center of mass plus a decoupled internal part, 
and can be mapped further into an isolated system using Niederer's transformation. 
Conversely, the manifest Galilean boost symmetry of the isolated system can be ``imported'' to the
oscillator and to the magnetic systems, respectively,
to yield the symmetry used by Gibbons and Pope to prove the separability.
For vanishing interaction potential the isolated system is free and our procedure endows all our systems with a hidden  Schr\"odinger 
symmetry, augmented with independent internal rotations.
All these properties follow from  the  cohomological  structure of the Galilei group, as explained by Souriau's  \emph{``d\'ecomposition barycentrique''}. 
\end{abstract}

\pacs{04.50.Cd,11.30.-j,02.20.Sv\\[8pt]
Physics Letter B (accepted)}


\maketitle

\section{Introduction}

Kohn's theorem \cite{Kohn}, commonly but vaguely ascribed  to
Galilean invariance,
says that  a system of charged particles in a uniform magnetic field can be
decomposed into center-of-mass and relative 
motion
if the charge/mass ratios are identical,
\beq
e_a/m_a=\epsilon=\const.\,
\label{Kcond}
\eeq
The term ``Galilean invariance'' has been 
 recently been criticized by Gibbons and Pope \cite{GiPo}, though, who argue that their symmetry transformation $\vx\to\vx+\va(t)$ is
 not of the usual Galilean form $\vx\to\vx+\vb t$,  
and belongs rather to the Newton-Hooke group. 
 
In this Note we show that the two, apparently contradictory, statements \emph{can} be conciliated~: 
$\va(t)$ \emph{is} a Galilean boost,
--- but it acts in a
way which is different form the usual one.
Separability \emph{does follow therefore from 
 ``abstract'' Galilean invariance --- as it does from Newton-Hooke symmetry also}. 
In detail, we show that when (\ref{Kcond}) holds the 
Bargmann space of the magnetic-background system is 
conformally related to an isolated system
 with ordinary boost symmetry, and
``importing'' it 
guarantees the existence of a rest frame also for the
magnetic-background. The ``imported boost"  coincides with the symmetry used by Gibbons and Pope \cite{GiPo}.

In the absence of an interaction potential, the system carries, moreover, a ``hidden'' Schr\"odinger symmetry obtained by ``importing'' that of a free system, augmented with internal rotations.
Our results
shed  new light on Kohn's theorem
and generalize Souriau's \emph{``d\'ecomposition barycentrique''} \cite{SSD}. 

\section{A ``relativistic'' proof of Kohn's theorem}

We demonstrate our statements in the Kaluza-Klein-type framework  \cite{Bargmann} 
which says that the  \emph{null geodesics of 
a manifold in $d+2$ dimensions} with Lorentz metric,
\begin{equation}
ds^2=
d\vx^2+2dtds-
\frac{2U}{m}(\vx,t)dt^2
\label{potmetric}
\end{equation}
 project,
for a particle in  $d+1$ dimensional non-relativistic space-time with coordinates  $(\vx,t)$, according to Newton's equations, $m\ddot{\vx}=-\vnabla U$. The generalization of (\ref{potmetric}) to $N$ particles in
$d$ dimensions in a potential $U$ is provided by
the $Nd+2$ dimensional metric \cite{Bargmann},
\beq
\sum_{a=1}^N\frac{m_a}md\vec{x}_a^2+2dtds-\frac{2U}{m}dt^2
\quad\hbox{where}\quad
m=\sum_{a=1}^Nm_a.
\label{Noscimet}
\eeq 

A remarkable property of the metric 
(\ref{potmetric}) is that it defines a preferential
Newton-Cartan structure \cite{DHNC} on non-relativistic spacetime 
obtained by projecting out the ``vertical'' direction generated by the lightlike vector $\p_s$
 \cite{Bargmann}.
In the quadratic case $U=\pm\half\omega^2\vx^2$, 
(\ref{potmetric}) describes, from 
the mechanical point of view,
 an [attractive of repulsive] harmonic oscillator \cite{Bargmann}.
In a relativistic language  (\ref{potmetric})
is a pp-wave, and 
the quotient is Newton-Hooke space-time
\cite{GiPo,DHNC}, which  carries  a
Newton-Hooke symmetry, represented by the isometries of the metric \cite{Bargmann,GiPo}.

But the metric (\ref{Noscimet})  is just one example of
a ``Bargmann'' spacetime, whose characteristic feature
is that it carries a covariantly constant lightlike vector \cite{Bargmann}. 
More generally, the metric can also accommodate a vector potential
 \cite{DHP2}~: the projections of the
null geodesics of
\begin{equation}
ds^2=
d\vx^2+2dt\big(ds+\frac{e}{m}{\vA}(\vx)\cdot d\vx\big)-
\frac{2e}{m}{U}(\vx)dt^2
\label{genmetric}
\end{equation}
satisfy the usual [Lorentz] equations of motion of a non-relativistic
particle in a (static) ``electromagnetic'' field
$
\overrightarrow{B}=\vnabla\times\overrightarrow{A},\,
\overrightarrow{E}=-\vnabla{U}.
$ 

A remarkable feature  is that,
in the plane, the
 isotropic oscillator metric (\ref{potmetric}) with $U=\half\omega^2\vx^2$  
 is indeed \emph{equivalent} to the ``magnetic'' metric (\ref{genmetric}) with  vector potential
 $A_i=-\frac{B}{2}\epsilon_{ij}x^j$, used to describe the motion in a uniform magnetic field  $B$ perpendicular to the plane \footnote{For simplicity, we took $B(t)=B=\const.$
 and work in the plane.}.
Switching to a rotating frame, 
\beq
\vX=\barray{c}X^1\\ X^2\earray=R_B\vx\equiv
\barray{cc}
\cos\Omega\, t &\sin\Omega\, t
\\[8pt]
-\sin\Omega\, t &\cos\Omega\, t
\earray
\barray{c}x^1\\ x^2
\earray;
\qquad
\Omega= \frac{eB}{2m}\,,
\label{Boscitr}
\eeq
completed with $T=t$ and $S=s$,
 carries the magnetic metric into that of the oscillator. (This is just the familiar Larmor trick in a new guise).
$N$ particles in the plane with electric charges
$e_a$ are described by adding to the metric (\ref{Noscimet})
$ 
2dt\sum_a(e_a/m)\vA_a\cdot d\vx_a,
$ 
where $A_a^i=-(B/2)\epsilon^i_{\ j}x_a^j$.

The generalization to $N$ particles being straightforward,  we restrict ourselves henceforth  to
two charged planar particles in a constant magnetic field. 
With the same choice of gauge for $\vA$ as above, we hence consider the $2\times2+1+1=6$-dimensional metric
\beq
\sum_a\frac{m_a}{m}\,d\vec{x}_a^2
+2dtds
-B\sum_a\frac{e_a}{m}\left(x_a^2dx_a^1-x_a^1dx_a^2\right)dt
-\frac{2V}{m}dt^2,
\label{2PB}
\eeq 
where we have included an interaction potential $V\equiv V(|\vx_a-\vx_b|)$ and dropped the external trapping potential $U$ for simplicity.
 Then, applying (\ref{Boscitr})
to each vector $\vx_a\; (a=1,2)$ yields
\begin{eqnarray*}
&
\sum_a\displaystyle\frac{m_a}md\vX_a^2+2dTdS-\displaystyle\frac{2V}{m}dT^2
+ \displaystyle\frac{\Omega}{m}\sum_a\left[\big(m_a\Omega-e_aB\big)\vX_a^2\right]dT^2
\\[8pt]
&+ \displaystyle\frac{1}{m}\sum_a\Big[\left(2
m_a\Omega-e_aB\right)\left(X_a^1dX_a^2-X_a^2dX_a^1\right)\Big]dT.
\end{eqnarray*}
Our clue is now that \emph{if the particles have the same charge to mass ratios}, (\ref{Kcond}),
 then, choosing the rotation frequency as
$ 
\Omega=\epsilon{B}/{2}
$ 
carries the constant-magnetic-field-metric,  (\ref{2PB}), into
\beq
\sum_a\frac{m_a}{m}d\vX_a^2+2dTdS-\frac{2}{m}\left(
\frac{\omega^2}{2}\sum_am_a\vX_a^2
+V\right)dT^2,
\qquad
\omega^2=\epsilon^2\frac{B^2}{4},
\label{aniosci}
\eeq
which is the metric for  an 
\emph{anisotropic} oscillator in $d=2+2$ dimensions, 
augmented with the potential $V$ \footnote{The relation of the
(non-commutative) Landau problem with an anisotropic harmonic oscillator has also been studied \cite{AGKP08}.}.
The two-particle metric (\ref{aniosci}) plainly decomposes into center-of-mass and relative coordinates.
Putting
\beq
\vX_0=\frac{m_1\vX_1+m_2\vX_2}{m},
\qquad
\vY=\frac{\sqrt{m_1m_2}}{m^2}(\vX_1-\vX_2)
\eeq
and calling
$V(|\vY|m^2(m_1m_2)^{-1/2})$ again $V(|\vY|)$
with some abuse of notations
(\ref{aniosci}) is indeed written as
\beq
\Big\{d\vX_0^2-\omega^2\vX_0^2dT^2
\Big\}+\Big\{d\vY^2-\big(
\omega^2\,\vY^2+\displaystyle\frac{2{V}(|\vY|)}{m}\big)dT^2\Big\}+2dTdS,
\label{decompmet}
\eeq
The first curly bracket here clearly describes
 the center-of-mass which behaves as
a planar particle of mass $m$ in an attractive oscillator
 field, to which the ``internal  vector'' $\vY$ adds two more dimensions, interpreted as an ``internal oscillator'' with an interaction potential.
Note that the ``external'' and ``internal'' oscillators have identical
frequencies $\omega$ and also that the anisotropic
oscillator became isotropic when expressed
in the new coordinates. The null geodesics of the metric 
(\ref{decompmet}) project to the decoupled 
 system of planar oscillators 
\beqa
\frac{d^2\vX_0}{dT^2}+\omega^2\vX_0=0,
\qquad
\frac{d^2\vec{Y}}{dT^2}+
\omega^2\vec{Y}+\frac{1}{m}\vnabla_Y{V}=0.
\label{decosceqmot}
\eeqa
The center-of-mass, $\vX$, performs  an elliptic 
``deferent'' motion around the origin,
to which $\vY$ adds an ``epicycle''
with the same oscillator frequency, plus some
internal interaction.
Transforming back to the magnetic background, we have
\beqa
\Big\{d\vx_0^2+\epsilon B(\vx_0\times d\vx_0)dt
\Big\}
+&\Big\{d\vy^2+\epsilon B(\vy\times d\vy\big)dt
-2\displaystyle\frac{{V}(|\vy|)}{m}dt^2\Big\}
+2dtds,
\label{Bdecompmet}
\\[6pt]
\ddot{x}_0^i=-2\omega\epsilon^{ij}\dot{x}_0^j,
\qquad
&\ddot{y}^i=-2\omega\epsilon_i^{\ j}\dot{y}^j
-\p_{y^i}{V}\,,
\label{decBeqmot} 
\eeqa
where ${\vx}_0=R_B^{-1}\vX$ is the magnetic center-of-mass
and ${\vy}=R_B^{-1}\vY$ is the internal coordinate.
\goodbreak

The decomposition (\ref{decompmet}) [or   
(\ref{decosceqmot})] allows us to infer that the system admits \emph{two independent and separately conserved angular momenta}, since one can 
consider \emph{independent external and internal rotations},
\beq
\barraynb{lll}
\vX_0\to R_{ext}\vX_0,\qquad &\vY\to\vY,
 &L_{0}= m\vX\times\dot{\vX},
\\[6pt]
\vX_0\to \vX_0, &\vY\to R_{int}\vY,\qquad
&L_{int}=m\vY\times\dot{\vY},
\earraynb
\label{2rot}
\eeq
where $R_{int}$ and $R_{ext}$ are planar rotation
matrices. The first  rotation corresponds to rotating the center of mass alone, and the second
corresponds to rotating it around its center of mass.
 The (separate) conservations of the
two angular momenta can be checked directly using
the equations of motion (\ref{decosceqmot}) or (\ref{decBeqmot}).
 
\section{Mapping to an isolated system and hidden Schr\"odinger symmetry}

Another remarkable feature of the metric 
(\ref{potmetric}) [with $\vx\leadsto\vX,\,
t\leadsto T$] is that, in the quadratic case
$2U=\pm\omega^2\vX^2$ and for uniform $B=B(t)$,  it is \emph{conformally flat} \cite{Bargmann,DHP2}. For $B=\const.$, lifting Niederer's transformation \cite{FreeOsc} to Bargmann space according to
\beq
\vXi=\frac{\vX}{\cos\omega T},
\quad
\tau=\frac{\tan\omega T}{\omega},
\quad
\Sigma=s-\frac{\omega}{2}\vX^2\tan\omega T 
\label{flattening}
\eeq
maps in fact each half-period of the oscillator
conformally into the free metric \cite{FreeOsc}, 
\hfill\break
$
d\vXi^2+2d\tau d\Sigma=\cos^{-2}\omega T
\big(d\vX^2+2dTdS-2UdT^2\big).
$ 
Generalizing (\ref{flattening}), 
we observe that
\beq
\vec{\Xi}_a=\frac{\vX_a}{\cos\omega T},
\qquad
\tau=\frac{\tan\omega T}\omega,
\qquad
\Sigma=s-\frac
{\omega}{2}\left(\sum_a\frac{m_a}{m}\vX_a^2\right)\tan \omega T
\label{Nflattening}
\eeq
carries the two-particle oscillator metric (\ref{aniosci}) into
\beq
\frac{1}{1+\omega^2\tau^2}\left(\sum_a\frac{m_a}{m}\,d\vXi_a^2+2d\tau d\Sigma-
2{V}^*d\tau^2\right)
\label{Nfreemetr}
\eeq
where ${V}^*={V}(|\vXi_1-\vXi_2|/\sqrt{(1+\omega^2\tau^2)})$,
which is conformal to an isolated system with some time-dependent interaction. The latter is plainly
decomposed into center-of-mass, $\vXi_0=\sum_am_a\vX_a/m$, and relative coordinate, 
$\vUpsilon=\sqrt{1+\omega^2\tau^2}\,(\vXi_1-\vXi_2)$,
as confirmed by writing (\ref{Nfreemetr}) as,
\beq
\frac{1}{1+\omega^2\tau^2}\Big(d\vXi_0^2+d\vUpsilon^2
+2d\tau d\Sigma
-\frac{2{V}^*}{m(1+\omega^2\tau^2)}d\tau^2\Big).
\eeq
Proceeding backwards, the c-o-m decompositions (\ref{decompmet}) and (\ref{Bdecompmet}) are recovered.

Note that the potential became, in general, time-dependent. A notable exception is the Calogero case
when the interaction potential is
a sum of inverse-squares, which is conformally invariant \cite{Calogero}.

Now the isolated system (\ref{Nfreemetr}) is clearly invariant under Galilei boosts acting in the ordinary way,
$
\vXi_a\to\vXi_a+\vbeta \tau,
\,
\tau\to\tau.
$
Boosts only act on the center-of-mass,
$\vXi_0$, and leave the internal coordinate, $\vUpsilon$, invariant.
Then the inverse of (\ref{Nflattening})
``exports'' these  boosts to the oscillator 
background, and combining  with the Larmor rotation (\ref{Boscitr})
backwards  transports everything to our original space while respecting the c-o-m decomposition, 
$
\vX_a\to
\vX_a+(\omega^{-1}\sin\omega t)\,\vbeta
$ and $\vx_a\to\vx_a+\va(t)$, respectively, where
\beq
\va(t)=(\omega^{-1}\sin\omega t)\,R_B^{-1}(t)\vbeta
\label{Bboost}
\eeq
The new, ``hidden'' boosts look quite
different from the usual Galilean expression;
being isometries, they also belong to the Newton-Hooke group, and  are in fact identical to those in \cite{GiPo}, as solution of
$\ddot{\va}=\epsilon\, \va\times B$.

All boosts only act on the center-of-mass and leave the relative coordinate invariant.
Their role can be understood as follows. 
The c-o-m of the isolated system, $\vXi_0$, moves as a free particle, i.e., with constant velocity.
It can be brought therefore to a halt by a suitable (ordinary) Galilei boost~: the massive system has a rest-frame.
 Transforming backwards through  (\ref{Nflattening})
 to the oscillator context provides us with the
 (elliptical) motion of the oscillator c-o-m
 $\vX_0$ --- whereas it also ``exports''
 the boost symmetry to the oscillator problem, yielding ``hidden boosts''. Applied to $\vX_0$, the latter still brings the
 c-o-m motion to a rest. 
Acting with  (\ref{Boscitr}) 
 backwards provides us, at last,
 with the c-o-m motion of the
original charge system in the uniform magnetic background.
(It is amusing to check that the rotation (\ref{Boscitr}) backwards  
converts the oscillator-{\it ellipses} into \emph{circles} in the magnetic problem, as it should.) All this is decoupled from the internal motions.

For $V\equiv0$ we have, in particular,
an isolated
system  of free particles for which the
boost symmetry plainly extends to (centrally extended) Schr\"odinger symmetry, identified with $\p_\Sigma$-preserving conformal transformations of the metric  \cite{Bargmann,DHP2}.
Then this $9$-parameter Schr\"odinger symmetry can be
``exported'' backwards~: the full system becomes ``hiddenly'' Schr\"odinger symmetric \cite{DHP2,Galaj}. 
 
Turning to the internal part, 
the second line in (\ref{2rot}) is obviously an ``internal'' symmetry which can be added to the ``external'' (i.e., center-of-mass) one to yield a 10 parameter symmetry group, isomorphic to
\beq
\hbox{(centrally extended-Schr\"odinger)}\times \SO(2)\label{hiddensymm}
\eeq
as symmetry of the oscillator \footnote{The system has in fact yet one more conserved quantity 
namely the ``internal energy'' $E_{int}$, related to the 
independent external and internal time translations \cite{SSD}.
This can be understood in a multiple-time framework \cite{DuKu}, which allows for independent internal and external time translations. Here we only consider $E_{int}=0$. }, and
combining with (\ref{Boscitr}) extends the statement to the magnetic problem~: as long as the charge-to-mass 
condition (\ref{Kcond}) holds, the $2$-charge
system carries the same ``hidden''  symmetry   (\ref{hiddensymm}) -- but realized in an  ``even more hidden way'' \cite{DHP2}.

\section{Souriau's ``d\'ecomposition barycentrique''}\label{Souriau}

Skipping technical details, we would like to mention that our results here  fit perfectly into Souriau's 
{\it ``d\'ecomposition barycentrique''} \cite{SSD}. Souriau argues in fact that having a rest frame and a corresponding center-of-mass decomposition only depend on the \emph{cohomological properties of 
the Galilei group, $G$,} \footnote{These same cohomological properties determine central extensions 
\cite{SSD}.}   and are independent of the concrete way 
Galilean symmetry is implemented.  

Consider in fact an arbitrary Galilei-invariant mechanical system in $d$ dimensions. Its  ``space of motions'' $M$
[Souriau's abstract substitute for the phase space]
is even dimensional. If its dimension is  the lowest possible one, namely $2d$, then the Galilei group  acts on it transitively, and the 
space of motions is a coadjoint orbit 
 endowed with the canonical symplectic structure
 of the
centrally extended Galilei group: it describes
 a free spinless particle. Souriau calls it an elementary system.

If the dimension of $M$ is
at least $2d+2$, then the action of the Galilei group is not more transitive; the system is not more elementary. Then Souriau proves
that, for \emph{non-vanishing mass}, $m\neq0$, $M$ is
 split into the direct 
product of the $2d$ dimensional Galilei coadjoint orbit
$M_0$ with another symplectic manifold,
$M_{int}$,
\beq
M=M_0\times M_{int}.
\label{Sdecom}
\eeq
$M_0$ describes the center of mass, and $M_{int}$
is characterized by the vanishing of all external Galilean conserved quantities~: it describes the ``internal'' motions in the rest frame.

Moreover, while the
Galilei group acts transitively and symplectically on $M_0$, its
subgroup composed of \emph{rotations and time translations}
acts  independently and also symplectically  on $M_{int}$  which carries hence an internal angular momentum and internal energy.  

\section{The quantum picture}

Our investigations, presented so far classically, also apply
in the quantum context. Remember first that a wave function,
$\psi$, is an equivariant function on Bargmann space,
$\p_s\psi=im\psi$, and, for a scalar particle, the quantum counterpart of motion along null-geodesics  is 
the massless Klein-Gordon equation \footnote{If the
Bargmann manifold is curved, conformal invariance 
requires adding
a curvature term to (\ref{KGeq})  \cite{Bargmann,DHP2}.} \cite{Bargmann,DHP2},
\beq
\dAlembert\psi=0,
\label{KGeq}
\eeq
where $\dAlembert$ is the Laplace-Beltrami operator associated with the Bargmann metric.
Equivariance then implies that 
$\Psi(\vx,t)=e^{ims}\psi(\vx,t,s)$ is a well defined ``ordinary'' wave function, for which (\ref{KGeq})
reduces to a Schr\"odinger equation. In the oscillator and magnetic cases we get,
\beqa
i\partial_T\Psi&=&\left[-\sum_a\left(\frac{1}{2m_a}\vnabla_a^2-\frac{1}{2}\omega
^2m_a\vX_a^2\right)+V(|\vX_1-\vX_2|)\right]\Psi,
\label{osciSch}
\\[6pt]
i\partial_t\Psi&=&\left[-\sum_a\frac{1}{2m_a}\left(\vnabla_a-ie_a\vA_a\right)^2+V(|\vx_1-\vx_2|)\right]\Psi,
\label{BSch}
\eeqa
respectively, as expected.
In c-o-m coordinates
 (\ref{decompmet}) and resp. (\ref{Bdecompmet}), one readily finds instead,
\beqa
i\p_T\Psi&=&\left[\left\{-\frac{\vnabla_{X_0}^2}{2m}
+\frac{m\omega^2}{2}\vX_0\strut{}^2\right\}
+
\left\{-\frac{\vnabla_{Y}^2}{2m}
+\frac{m\omega^2}{2}\vY\, \strut{}^2
+{V}(|\vY|)\right\}
\right]\Psi
\label{osciSch2}
\\[8pt]
i\p_t\Psi&=&\left[\left\{
-\frac{1}{2m}(\vnabla_{x_0}-iq{\vA}_{\vx_0})^2\right\} 
+\left\{-\frac{1}{2m}\left(\vnabla_{y}-iq{\vA}_{%
\vy}\right)^2+V(|\vy|)\right\}
\right]\Psi,\quad
\label{BSch2}
\eeqa
where 
$q=\sum_a e_a$ is the total electric charge. Both equations are plainly separable
in c-o-m and internal coordinates.

The implementation of symmetries on wave functions
can also be deduced from the Bargmann picture \cite{Bargmann,DHP2}~:  a conformal transformation $f$
of Bargmann space with conformal factor $\Omega_f^2$,
 acts as 
$ 
\widehat{f}\,\psi=\Omega_ff^*\psi.
$ 
 If  $f$  also preserves the lightlike vector $\p_s$, then
it projects into a transformation
$F(\vx,t)$ of non-relativistic spacetime, and locally
$f(\vx,t,s)=\big(F(\vx,t),s+\sigma(\vx,t)\big)$. 
Our symmetry is implemented therefore on an ordinary wave function according to
\beq
\widehat{F}\,\Psi(\vx,t)=\Omega_f(\vx,t)e^{i\sigma(\vx,t)}\Psi(F(\vx,t)).
\label{downact}
\eeq
For our three types of boosts 
with parameter $\vbeta$ we get, in particular,
\beqa
&e^{-i\vbeta\cdot\sum_am_a\vXi_a-m\frac{\vbeta^2}{2}\tau}\Psi(\vXi_a',\tau),
\qquad
&\vXi_a'=\vXi_a+\vbeta\tau
\label{Qboostfree}
\\[8pt]
&
e^{-i\cos\omega T\,\vbeta\cdot\sum_am_a\vX_a-
m\frac{\vbeta{}^2}{2}\frac{\sin\omega T}{\omega}
}
\Psi(\vX_a',T),
&\vX_a'=\vX_a+(\omega^{-1}\sin\omega T)\,\vbeta,
\quad
\label{Qboostosc}
\\[8pt]
&e^{-i\cos\omega t\,R_B^{-1}(t)\vbeta\cdot\sum_am_a\vx_a-
m\frac{\vbeta{}^2}{2}\frac{\sin\omega t}{\omega}
}
\Psi(\vx',t),
&\vx_a'=\vx_a+(\omega^{-1}\sin\omega t)\,R_B^{-1}(t)\vbeta
\label{QboostB}
\eeqa
Note that (\ref{Qboostosc}) reduces to (\ref{Qboostfree})
as $\omega\to0$.

\section{Conclusion}

We have proved that
the decomposition into center-of-mass and internal
motion \emph{is} indeed a consequence of
Galilean symmetry as it is popularly said --- 
but boosts act in a ``hidden'', and 
 not in the conventional way.
As the Newton-Hooke group has a similar of cohomological structure as the Galilei group \cite{Negro}, this implies the separability, (\ref{decompmet}) and
(\ref{decosceqmot}), of a particle system in an
oscillator background directly from the group theory,
applied to the Newton-Hooke group. Details will be presented elsewhere \cite{TBP}.

Souriau's theorem also explains why we do not have
similar properties in the relativistic case: the
 Poincar\'e group has trivial cohomology.

\begin{acknowledgments} 
P.A.H is indebted to the \textit{
Institute of Modern Physics} of the Lanzhou branch of
the Chinese Academy of Sciences for hospitality.
We would like to thank C. Duval for illuminating discussions and A. Galajinsky for correspondence.
This work has been partially supported by the National Natural Science Foundation of 
China (Grant No. 11035006) and by the Chinese Academy of Sciences visiting 
professorship for senior international scientists (Grant No. 2010TIJ06). 
\end{acknowledgments}
\goodbreak


\end{document}